\documentclass[prl,twocolumn,showpacs,preprintnumbers,superscriptaddress,
amsmath,amssymb,tightenlines,epsfig]{revtex4}

\usepackage{bm}% bold math
\usepackage{graphicx}% Include figure files
\usepackage{graphicx}% Include figure files
\usepackage{dcolumn}% Align table columns on decimal point
\usepackage{bm}% bold math

\newcommand{\ket}[1]{|#1\rangle}

\newcommand{\beq}{\begin{equation}}
\newcommand{\eeq}{\end{equation}}

\newcommand{\bml}{\begin{mathletters}}
\newcommand{\eml}{\end{mathletters}}
\newcommand{\commentout}[1]{{}}

\newcommand{\rv}{{\bf r}}

\newcommand{\dv}{{\bf d}}
\newcommand{\Ev}{{\bf E}}
\newcommand{\Pv}{{\bf P}}

\normalsize

\begin{document}

\title{Transfer and storage of vortex states in light and matter waves.}

\author{Zachary Dutton}
\affiliation{National Institute of Standards Technology,
Electron and Optical Division, Gaithersburg MD 20899-8410}
\author{Janne Ruostekoski}
\affiliation{Department of Physics, Astronomy and Mathematics, University of
Hertfordshire, Hatfield, Herts, AL10 9AB, UK}
\date{\today}

\begin{abstract}

We theoretically explore the transfer of vortex states between
atomic Bose-Einstein condensates and optical pulses using
ultra-slow and stopped light techniques. We find shining a
coupling laser on a rotating two-component ground state condensate
with a vortex lattice generates a probe laser field with optical
vortices. We also find that optical vortex states can be robustly
stored in the atomic superfluids for long times ($>100~$ms).
\end{abstract}


\maketitle

One of the most dramatic developments in recent experiments on
atomic Bose-Einstein condensates (BECs) is the creation and study
of quantized vortices and vortex lattices
\cite{ANG02,twoCompVortex}. In this Letter we show that light
modes with orbital angular momentum can be efficiently stored in
superfluid vortex configurations and later rewritten back onto
light fields. This also provides a feasible mechanism to transfer
BEC vortex lattice ground states to light pulses. The strong
light-matter coupling can be implemented using the recent
technology of ultra-slow and stopped light pulses in atomic vapors
\cite{Nature1,Nature2,OtherStoppedLight,stoppedLightTheory}, based
on the method of electromagnetically-induced transparency (EIT)
\cite{EIT}. In these experiments, an optical field's amplitude and
phase pattern was written onto superpositions of atomic states,
stored for several milliseconds, and then re-written back onto the
original light field and output.

One of the unique characteristics of slow light is the strong yet
coherent action of the matter fields on the light fields
\cite{SolitonVortex}. We here study the robust storage and
transfer of coherent information in terms of vortex
configurations, for experimentally feasible parameters of a
two-component $^{87}$Rb BEC. We find that shining a coupling laser
pulse can generate a probe pulse which contains optical vortices
corresponding to the vortices originally in the atomic fields. If
vortices are present in both BECs, the vortex circulation of one
BEC component is reversed in the probe field. This light field
could potentially be input into a second BEC which is optically
connected with the first BEC (even if it is spatially distant),
allowing transfer of vortex states between BECs.

\begin{figure}
\includegraphics[width=0.9\columnwidth]{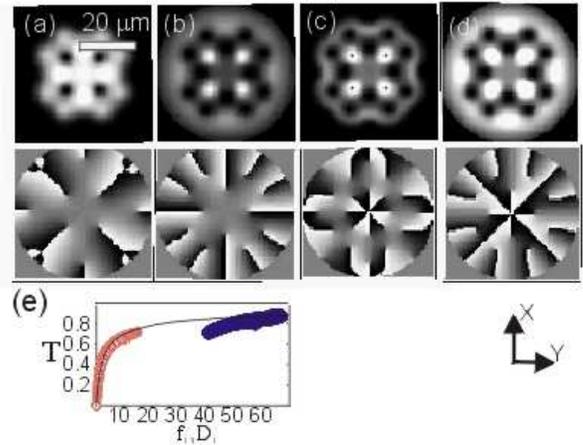}
\caption{\label{fig:lattice}\textbf{Writing an atomic vortex
lattice onto a light field.} The optical density (top row) $D_i =
N \sigma \int dz |\psi_i|^2$ (black=0, white=55) and the phase
$\phi_i$ in the $z=0$ plane (bottom row) for a vortex lattice
\textbf{(a)} in $\ket{1}$ and \textbf{(b)} in $\ket{2}$.
\textbf{(c)} $n_p$ (the number of probe photons output per cross
sectional area $\sigma$) upon a switch-on of the coupling field
(black=0, white=25) and the phase profile of $\phi_p$, which
contains vortices of both circulation. \textbf{(d)} Same
quantities as (c) for a non-rotating ground state BEC in
$\ket{1}$, resulting in vortices of only one circulation.
\textbf{(e)}  A scatter plot of the transfer efficiency
$T=n_p/n_2$, where $n_2$ is the number of $\ket{2}$ atoms per
$\sigma$ before the switch-on, versus $f_{13} D_1$ for the case in
(c) (open circles, red online) and for (d) (closed circles, blue
online). The solid curve shows the estimate $1-(1 +
f_{13}D_1)^{-1/2}$. }
\end{figure}

The strong coupling between the quantized optical and BEC vortices
is an interesting phenomenon in its own right, as a demonstration
of nonlinear superfluid-light optics and may be useful in imaging
vortex lattices.  But it also points to possibilities for
information processing and storage. Experiments have seen advances
in the generation of Laguerre-Gaussian (LG) modes of light with
orbital angular momentum as well as applications of these modes to
quantum information \cite{Vaziri}. It may be possible to utilize
modes with several different angular momentum quantum numbers to
build a quantum information architecture based on a larger
alphabet than the traditional two-state systems
\cite{largeAlphabets}.  However, optically based quantum
information schemes suffer from the difficulty of trapping and
storing optical fields.

Here we perform a comprehensive analysis of how LG beams can be
written into BECs, stored in the atomic fields, and later
rewritten onto light fields.  We find one can choose parameters so
the coherent BEC dynamics do not dissipate the information during
the storage time, allowing storage fidelities on the order 70\%
for several hundred milliseconds with typical parameters. This
fidelity can be further improved by using BECs with larger optical
densities. Moreover, the phase information in vortex lines can
exhibit very long life times of tens of seconds or can even be
energetically stable \cite{ANG02}. While similar LG beams have
been proposed as a method to generate vortices in BECs \cite{LG}
and in experiments vortices have been imprinted on BECs
\cite{twoCompVortex}, our scheme here is very different: In the
nonlinear light-matter coupling both amplitude and phase are
robustly and simultaneously exchanged, allowing the controlled
storage of information. Moreover, slow light and EIT are now
inspiring important coherent matter wave technology applications,
e.g., BEC shock waves \cite{SolitonVortex}, probing decoherence in
BECs \cite{RUO99b}, optical black holes \cite{ulf}, storage and
production of non-classical states \cite{quantumStorage,squeezed},
quantum information processing \cite{quantumInfo}, low light level
nonlinear optics \cite{switch}, and optical processing
\cite{processing}.

We consider two BECs of Rb-87 atoms, in stable internal states
$\ket{1}\equiv\ket{5S_{1/2},F=2,M_F=+1}$ and
$\ket{2}\equiv\ket{5S_{1/2},F=1,M_F=-1}$, with macroscopic
wavefunctions $\psi_1$ and $\psi_2$.  These are connected, by
resonant, $+z$ propagating {\it coupling} and {\it probe} light
fields, $\Omega_c$ and $\Omega_p$ (each of wavelength
$\lambda=780$~nm), to the excited state
$\ket{3}\equiv\ket{5P_{1/2},F=2,M_F=0}$, which decays at
$\Gamma=(2\pi)~6~\mathrm{MHz}$, forming a $\Lambda$ three-level
structure. Here the Rabi frequencies $\Omega_{i}(\rv,t) \equiv
-\dv_{i3}\cdot\Ev_{i}(\rv,t)$ ($p=1,c=2$) are defined in terms of
the atomic dipole matrix elements $\dv_{i3}$ and the slowly
varying envelope (SVE) of the electric fields $\Ev_{i}$ (with the
rapid phase rotation at the optical frequencies and optical
wavenumbers factored out). The light propagation then follows from
SVE approximation to the Maxwell's equations \cite{diffraction}:
\beq \label{eq:formalismP} \left( \frac{\partial}{\partial z}+
\frac{1}{c}\frac{\partial}{\partial t} \right) \Omega_{i}  =
-{ik\over \hbar\epsilon_0}\,\dv_{i3}\cdot\Pv_{i3}\,. \eeq
\noindent Here the SVE of the polarization is
$\dv_{i3}\cdot\Pv_{i3}\simeq f_{i3}{\cal D}^2 N\psi^*_i\psi_3$,
$N$ is the initial total number of BEC atoms, $f_{13}=1/4$ and
$f_{23}=1/12$ are dimensionless oscillator strengths, ${\cal
D}=\sqrt{\hbar \epsilon_0 \sigma \Gamma/k}$ is the reduced dipole
matrix element and $\sigma\equiv 3 \lambda^2/2 \pi$ is the
resonant cross-section. The BEC wavefunctions $\psi_1,\psi_2$
evolve according to generalized Gross-Pitaevskii equations (GPEs):
\beq \label{eq:formalism1} i \hbar  \dot\psi_{i} = \big(H_0+
\sum_{j=1,2} U_{ij} | \psi_j |^2 \big)
\psi_{i}+\hbar\Omega^*_i\psi_3\,, \eeq \noindent where $H_0\equiv
-\hbar^2\nabla^2/2m + V$, with a perfectly overlapping trapping
potential for both states $V(\mathbf{r}) =\frac{m}{2} [\omega_r^2
(x^2+y^2)+\omega_z^2 z^2]$. In the interaction coefficients
$U_{ij} = 4 \pi N \hbar^2 a_{ij}/m$, the $a_{ij}$ are the
scattering lengths between atoms in states $|i \rangle$ and
$|j\rangle$. For $^{87}$Rb, $a_{12}=5.5$~nm and
$a_{11}:a_{12}:a_{22}::0.97:1.0:1.03$ \cite{scatteringLengths}.
Often in practice, and throughout this Letter, we will be working
in a regime where, whenever the light fields are present, the
atomic internal state dynamics and light field couplings are on a
much faster time scale and dominate the external dynamics.  In the
context of slow light experiments we can adiabatically eliminate
$\psi_3\simeq - i (\Omega_{p} \psi_{1}+\Omega_{c}\psi_{2}
)/\Gamma$ in Eqs.~(\ref{eq:formalismP}-\ref{eq:formalism1})
\cite{processing}. Then the last term in Eq.~(\ref{eq:formalism1})
results in both coherent exchange between $|1\rangle,|2\rangle$ as
well as absorption into $|3 \rangle$. In our model, atoms which
populate $|3 \rangle$ and then spontaneously emit are assumed to
be lost from the BECs.

Vortex lattices can be produced by imparting angular momentum to
the system with a time-dependent rotating potential. We
numerically find the ground state of a two-component BEC rotating
along the $z$ axis at the speed $\bar\omega=0.3 \, \omega_r$ by
evolving the GPEs (\ref{eq:formalism1}) in imaginary time in the
absence of the light fields in the rotating frame, obtained by
replacing $H_0$ by $H_0-\bar\omega \hat{L}_z$, with
$\hat{L}_z=i\hbar(y\partial_x-x\partial_y)$. At every time step,
we separately normalized the wavefunctions to fix the atom number
in each component. The full 3D integration, without imposing any
symmetry on the solution, is performed in a `pancake-shaped' trap
$\omega_r=0.1 \omega_z=(2 \pi)~10~\mathrm{Hz}$, with $N=2.76
\times 10^5$, on a spatial grid of $256^2\times32$.

In Fig.~\ref{fig:lattice} we present a calculation of such a
two-species lattice. The phases of the wavefunctions $\phi_i$
indicate a lattice of singly-quantized vortices in each component,
with the positions of the vortex cores in the two species offset
from each other. The filling of the vortex cores by the other BEC
significantly increases the core size, as compared to vortices in
a single-component BEC, and makes them observable even without a
ballistic expansion. The 3D results are qualitatively similar to
the previous 2D calculations \cite{MUE}: The rectangular vortex
lattice pattern is recognizable close to the trap center, but
becomes distorted close to the BEC boundaries. The total density
in the center is $6.9 \times 10^{13} \mathrm{cm}^{-3}$. The vortex
core positions (and $\phi_1,\phi_2$) change very little along $z$.

To model the writing of this vortex lattice to the probe field
$\Omega_p$ we numerically solved
Eqs.~(\ref{eq:formalismP}-\ref{eq:formalism1}) when a coupling
field with a peak input value of $\Omega_c^{(\mathrm{in})}=(2
\pi)~4~\mathrm{MHz}$ was switched on suddenly (in a time $0.8~\mu
\mathrm{s}$).  When this happens the coherence $\psi_1^\ast
\psi_2$ acts to generate a probe field $\Omega_p$ in such a way
that the contributions to the polarization in
Eq.~(\ref{eq:formalismP}) cancel out, and a dark state is formed.
The resulting output intensity and phase pattern of $\Omega_p$ is
shown in Fig.~\ref{fig:lattice}(c).

To understand the results, we note that solving
Eq.~(\ref{eq:formalismP}), ignoring the negligble $z$ variation of
$\psi_1/\psi_2$, shows the fields eventually reach the asymptotic
values $\Omega_p=-\Omega_c^{(\mathrm{in})} (f_{13} \psi_1^\ast
\psi_2)/(f_{13} |\psi_1|^2+f_{23} |\psi_2|^2)$ and
$\Omega_c=\Omega_p\psi_1/\psi_2$. Note that $\Omega_p \rightarrow
0$ whenever {\it either} $\psi_1,\psi_2 \rightarrow 0$, giving
vanishing intensity at all the phase singularities. The phase
$\phi_p$ is determined by the relative BEC phase
$\phi_p=\phi_2-\phi_1+\pi$ (the coupling field acquires no phase
shift here).  Thus $\phi_p$ contains singularities with one
circulation at the location of vortices in $\psi_2$ and of the
opposite circulation at vortices in $\psi_1$.

In the columns where $|\psi_2/\psi_1| \ll 1$ the coupling field
coherently generates the probe field via coherent transfers from
$\ket{2}$ and $\ket{1}$ so that the number of probe photons output
per area $\sigma$, $n_p$, equals the original number of $\ket{2}$
atoms per $\sigma$, $n_2$. Deviations from this ideal limit are
due to absorptions in a `preparation' region (equal to one optical
depth) before the dark state is established.   When the optical
depth of $\ket{1}$, $f_{13} D_1 \gg 1$, this loss is relatively
small. On the other hand, near the vortex cores of $\ket{1}$ where
$\psi_1 \rightarrow 0$ there is not enough coherence to generate
$n_2$ photons. Instead $\Omega_c$ is attenuated by absorption and
eventually the dark state is established primarily via incoherent
spontaneous emission events. We found that the transfer efficiency
$T=n_p/n_2$ for each column can be determined by the optical
depth, agreeing very well with the prediction
$1-1/\sqrt{f_{13}D_1}$ which approaches unity as $D_1 \rightarrow
\infty$; see Fig.~\ref{fig:lattice}(e). The overall output
efficiency integrated over $x,y$ is 35\%.

Motivated by the fact that the fidelity improves with $D_1$ we
also considered coupling the $\psi_2$ vortex lattice of
Fig.~\ref{fig:lattice}(b) to a vortex-free, non-rotating ground
state BEC in $\ket{1}$, with $N=3 \times 10^6$
[Fig.~\ref{fig:lattice}(d)].  There the peak optical density
$f_{13} D_1$ is $70$ and is $>35$ throughout the region occupied
by $\ket{2}$. To a good approximation, the probe intensity simply
reflects $\psi_2$ [see Fig.~\ref{fig:lattice}(b)] and only
contains vortices of one helicity. The fidelity is now higher
[Fig.~\ref{fig:lattice}(e)] with the overall efficiency 85\%.

\begin{figure}
\includegraphics[width=0.9\columnwidth]{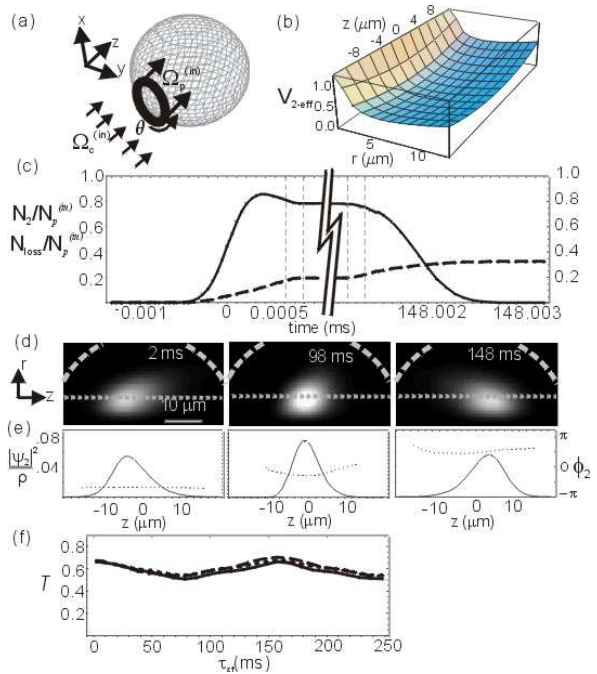}
\caption{\label{fig:example}\textbf{Inputting, storing, and
outputting an optical vortex in a BEC.} \textbf{(a)} A spherical
BEC (grey sphere) is originally illuminated by a c.w. coupling
field $\Omega_c$.  A probe field pulse in an LG (`doughnut') mode
(black torus), is then input and stopped when it has propagated to
the center. \textbf{(b)} The effective potential seen by $\psi_2$,
$V_{2-\mathrm{eff}}(\mathbf{r})$ in units of $\hbar \omega_z$,
during the storage for $m=1$, with a minimum near $r=6\mu$m,
$z=0$. \textbf{(c)} The number of atoms in $\ket{2}$, $N_2$,
relative to the number of input photons $N_p^{\mathrm{(in)}}$
(solid curve) and the number of atoms lost due to spontaneous
emission (dashed curve) during the probe pulse input and coupling
switch-off (the vertical dotted lines shown the points at which
the coupling field is 90\% and 10\% of its full value). These are
constant during the storage time (the time break). After the time
break we then plot the same quantities during the coupling
switch-on and probe pulse output. \textbf{(d)} The relative
density profile $|\psi_2|^2/\rho$ during the storage at the times
indicated. The dotted curve indicates the BEC boundary. The dotted
lines shows the cuts plotted in (d). \textbf{(e)}
$|\psi_2|^2/\rho$ and phase of $\phi_2$ at the same lines along
the cuts indicated in (c). \textbf{(f)} The corresponding output
fidelity $\mathcal{T}$ upon switch-ons after various storage times
$\tau_{\mathbf{st}}$, for $m=1,2,0$ (solid, dot-dashed, and dotted
lines, respectively).}
\end{figure}

Having determined that vortex states can indeed be written from
atoms to light fields we now consider the possibility of storing
information, input in the form of vortices in light fields, in a
two-species BEC.  For this study, we switch the roles of the two
stable $^{87}$Rb states so $\ket{3}$ is coupled to
$\ket{2}\equiv\ket{F=2,M_F=+1}$ by $\Omega_c$ and to
$\ket{1}\equiv\ket{1,-1}$ by $\Omega_p$. A BEC is initially in the
(non-rotating) ground state of $\ket{1}$, labelled
$\psi_1^{(\mathrm{G})}(\mathbf{r})$, in an isotropic trap
$\omega_z=\omega_r=(2 \pi)21~\mathrm{Hz}$, with $N = 4\times
10^6$, $R_{\mathrm{TF}}=25~\mu\mathrm{m}$, peak density of $1.5
\times 10^{14}~\mathrm{cm}^{-3}$, and a peak optical density of
$f_{13} D_1=213$. The initial coupling field is at
$\Omega_c^{\mathrm{(\mathrm{in})}}=(2 \pi)~8~\mathrm{MHz}$ [see
Fig.~\ref{fig:example}(a)], while the LG probe input reads (in
cylindrical coordinates \cite{cylindar}):
\beq
\label{eq:laguerre}
\Omega_p^{(\mathrm{in})}(r,\theta,t)=\Omega_p^{(0)}
\bigg(\frac{r\sqrt{2}}{w}\bigg)^{|m|} e^{-\frac{r^2}{w^2}}e^{i m
\theta} e^{-\frac{t^2}{2 \tau_0^2}}\,,
\eeq
\noindent where $w$ is
the beam waist size. Each probe photon in the LG mode with $|m|\ge
1$ contains a vortex at $r=0$ with a vanishing intensity and $m$
units of orbital angular momentum.

We consider an $m=1$ mode with $\tau_0=0.25~\mu\mathrm{s}$,
$w=6.3~\mu\mathrm{m}$ and $\Omega_p^{(0)}=(2 \pi)~4~\mathrm{MHz}$.
This pulse contains $N_p^{\mathrm{(in)}}=3.83 \times 10^4$ photons
and propagates slowly, with little attenuation or distortion,
through the BEC, with the group velocity
$v_g\propto\Omega_c^2/|\psi_1|^2$. As it is input, the pulse
induces coherent transfer of atoms from $\ket{1}$ to $\ket{2}$.
Once the pulse is completely input ($0.6~\mu$s), $N_2$ nearly
reaches $N_p^{\mathrm{(in)}}$; see Fig.~\ref{fig:example}(c).
Because of the finite bandwidth of the EIT transmission window
there is some non-adiabatic loss from the BEC during the
propagation, due to absorption, and so the number of spontaneous
emission events $N_{\mathrm{loss}}$ reaches about $0.21
N_p^{\mathrm({in})}$ at this point.

This pulse is sufficiently weak $|\Omega_p| \ll
|\Omega_c^{(\mathrm{in})}|$ that the coupling field $\Omega_c$ is
nearly homogenous across the BEC and $\psi_1\simeq
\psi_1^{(\mathrm{G})}$ during the propagation.  Then $\psi_2$ is
coherently driven into the dark state
$\psi_2=-\psi_1^{(\mathrm{G})}(\Omega_p/\Omega_c^{(\mathrm{in})})$
and so acquires the amplitude and vortex phase pattern of the
input LG mode.  At $t=0.6~\mu$s, $\Omega_c$ is switched off in
about $0.2~\mu$s, and $N_2$ and $N_\mathrm{loss}$ are unaffected
by this rapid switch-off; see Fig.~\ref{fig:example}(c). Both
$\Omega_c$ and $\Omega_p$ smoothly ramp to zero intensity and the
LG mode is then stored in $\psi_2$. Figures~\ref{fig:example}(d-e)
show the relative density profiles $|\psi_2|^2/\rho$, where
$\rho\equiv|\psi_1|^2+|\psi_2|^2$ (note the density is
cylindrically symmetric \cite{cylindar}), and cuts of the density
and phase. The slow light propagation introduces virtually no
phase gradient along $z$ or $r$.

For times short compared to the BEC dynamics, determined by the
GPEs~(\ref{eq:formalism1}) with $\Omega_i=0$ (typically
milliseconds), any spatial mode can then be robustly stored.
However, much longer storage can be achieved if the pulse
parameters are chosen such that $\psi_1$ and $\psi_2$ remain
nearly stationary. In the weak probe limit $\psi_1\simeq
\psi_1^{(\mathrm{G})}$, while $\psi_2$ will evolve according to an
effective potential $V_{2-\mathrm{eff}}(\mathbf{r})$, determined
by the sum of the centrifugal potential $m^2/r^2$, the trap
$V(\mathbf{r})$ and the mean field potential $U_{12} |\psi_1|^2$;
see Fig.~\ref{fig:example}(b). Our choice of width $w$ and length
(which is governed by the pulse length $\tau_0$ and group
velocity), are such that $\psi_2$ nearly matches the ground state
of the potential $V_{2-\mathrm{eff}}$, which is trapping and
approximately harmonic for $a_{11}>a_{12}$. This results in the
dynamics being greatly suppressed, as seen in the second and third
panels of Figs.~\ref{fig:example}(d-e). The dominant motion is a
residual dipole sloshing in $z$, an unavoidable artifact of the
group velocity and pulse length having a small dependence on $r$.
For our parameters $148~\mathrm{ms}\simeq$ the half-period of
$V_{2-\mathrm{eff}}$, so $\psi_2$ at this time is nearly the
mirror image of $\psi_2$ at the switch-off time.

After an arbitrary time, one can then switch the coupling field
back on, regenerating a probe pulse according to
$\Omega_p=-\Omega_c^{(\mathrm{in})}(\psi_2/\psi_1^{(\mathrm{G})})$
\cite{processing} which is then output. Fig.~\ref{fig:example}(c)
shows $N_2$ and $N_{\mathrm{loss}}$ upon a switch-on at 148~ms.
The total accumulated of losses after the pulse has been output is
$0.33 N_p^{\mathrm{(in)}}$ and so $N_p^{\mathrm{(out)}}=0.67
N_p^{\mathrm{(in)}}$ photons are output, nearly identical to the
63\% transmission we would expect if one had propagated the LG
mode through the BEC without stopping and storing it. Furthermore,
the vortex phase profile has been preserved and little other
unwanted phase gradients are introduced.

We performed a series of numerical calculations of output pulses
for the case in Fig.~\ref{fig:example}, but with switch-ons at
various times. Since, ultimately, useful information storage
requires an alphabet with several different angular momentum
states, we also varied $m$. The dynamics are sufficiently
suppressed and the variation of the storage efficiency
$\mathcal{T}=N_p^{\mathrm{(out)}}/N_p^{\mathrm{(in)}}$ with
storage time is not significant; see Fig.~\ref{fig:example}(f).
The variations of $\mathcal{T}$ can be accurately estimated by the
model of \cite{processing}. They are due to spatial features
generated in the BECs during the dynamics which get written into
small temporal features on the revived probe fields, resulting in
additional EIT bandwidth loss. We will report more details of
these studies elsewhere.

The general utility of this method for storing fields with a
non-uniform phase profile depends not only on the number of
photons output but also on maintaining the phase pattern during
the storage. However, if, e.g., the vortex is pinned down by an
external potential \cite{laserPotentials}, the topological
stability of the vortices in BECs makes the winding number $m$
very robust, even in the presence of substantial unwanted phase
gradients.

We have shown how vortices can be written from atomic to optical
fields and vice-versa, providing a unique nonlinear technique to
create optical vortices as well as directly image phase
singularities in BECs. This method should also allow the transfer
of vortex configurations between BECs which are optically
connected.  Furthermore, we considered the ability of a BEC to
robustly store a vortex state input from light fields. While the
case studied here allowed 70\% efficiency, this number can be
further improved by using larger BECs (with the loss $\propto
1/\sqrt{D_1}$).  We have found that an appropriate choice of
parameters can suppress the BEC dynamics.  The ultimate storage
times should then only be limited by decoherence due to the small
inelastic collision rates and thermal and quantum fluctuations
($>100~$ms) \cite{scatteringLengths}.

\acknowledgments{We acknowledge financial support from the EPSRC.}


\begin{thebibliography}{99}
\bibitem{diffraction} Throughout this paper we ignore diffraction
effects which would be important when the transverse features in
the BEC (the vortex core size, etc.) become comparable $\lambda$.
\bibitem{cylindar} Because of the cylindrical symmetry we do not explicity calculate the dynamics of $\theta$ but
simply assume $\psi_2,\Omega_p \propto e^{i m
\theta}$.  Consequently, the stability of the vortex states
against drift from the trap center is not considered.
\end{thebibliography}
\end{document}